\documentclass[letterpaper]{article} %
\usepackage[margin=1.2in]{geometry}
\usepackage{xspace, graphicx, color, todonotes}
\usepackage{algorithm, balance}
\usepackage[noend]{algorithmic}

\newcommand{\emd}{ExaMiniMD\xspace}
\newcommand{\simgrid}{SimGrid\xspace}
\newcommand{\simsitu}{\textsc{Sim-Situ}\xspace}
\newcommand{\ie}{i.e.,\xspace}
\newcommand{\eg}{e.g.,\xspace}
\newcommand{\is}{\emph{in-situ}\xspace}
\newcommand{\intransit}{\emph{in-transit}\xspace}
\newcommand{\smpirun}{\texttt{smpirun}\xspace}
\newcommand{\smpicc}{\texttt{smpicc}\xspace}

\newcommand{\mpirun}{\texttt{mpirun}\xspace}
\newcommand{\mpicc}{\texttt{mpicc}\xspace}

\newcommand{\ema}[1]{\ensuremath{#1}\xspace}
\newcommand{\stride}{\ema{T}}
\newcommand{\coreRatio}{\ema{R}}

\newcommand{\simu}[1]{\ema{S_{#1}}}
\newcommand{\ana}[1]{\ema{A_{#1}}}
\newcommand{\ingest}[1]{\ema{\text{Ing}_{#1}}}
\newcommand{\get}[1]{\ema{R_{#1}}}
\newcommand{\collect}[1]{\ema{C_{#1}}}
\newcommand{\send}[1]{\ema{W_{#1}}}
\newcommand{\iter}{\ema{N}}
\newcommand{\totalana}{\ema{\rho}}
\newcommand{\idle}[2]{\ema{I_{#1}^{#2}}}
\newcommand{\makespan}[1]{\ema{m_{#1}}}
\newcommand{\efficiency}{\ema{\eta}}
\newcommand{\idleS}{IS\xspace}
\newcommand{\idleA}{IA\xspace}
\newcommand{\pp}[1]{\medskip\noindent\textbf{\emph{#1.}}\xspace}

\AtBeginDocument{
  \definecolor{pdfurlcolor}{rgb}{0,0,0.6}
  \definecolor{pdfcitecolor}{rgb}{0,0.6,0}
  \definecolor{pdflinkcolor}{rgb}{0.6,0,0}
  \definecolor{light}{gray}{.85}
  \definecolor{vlight}{gray}{.95}
}
\usepackage{url} 
\usepackage[colorlinks=true,citecolor=pdfcitecolor,urlcolor=pdfurlcolor,linkcolor=pdflinkcolor,pdfborder={0 0 0}]{hyperref}
\usepackage{hyperref}
\usepackage{amsmath}
\usepackage{cite}
\usepackage[caption=false, font=footnotesize]{subfig}
\usepackage{tikz}
\usetikzlibrary{patterns, calc}

\title{\simsitu: A Framework for the Faithful Simulation of \is
  Workflows}

\author{
   Valentin Honoré$^1$,
   Tu Mai Anh Do$^2$,
   Loïc Pottier$^2$,\\
   Rafael Ferreira da Silva$^3$,
   Ewa Deelman$^2$,
   Frédéric Suter$^1$
}
 \date{
\small{
    $^1$IN2P3 Computing Center / CNRS,   Lyon - Villeurbanne, France\\
    \it{valentin.honore@cc.in2p3.fr}, \it{fsuter@cc.in2p3.fr}\\
    $^2$USC Information Sciences Institute, Marina del Rey, CA, USA\\
    \it{tudo@isi.edu}, \it{lpottier@isi.edu}, \it{deelman@isi.edu}\\
    $^3$Oak Ridge National Laboratory, Oak Ridge, TN, USA\\
    \it{silvarf@ornl.gov}
}
}

\begin{document}

 \clearpage\maketitle \thispagestyle{empty}

\bigskip

\bigskip

\begin{abstract}
  The amount of data generated by numerical simulations in various scientific
domains such as molecular dynamics, climate modeling, biology, or astrophysics,
led to a fundamental redesign of application workflows. The throughput and the
capacity of storage subsystems have not evolved as fast as the computing power
in extreme-scale supercomputers. As a result, the classical post-hoc analysis
of simulation outputs became highly inefficient. In-situ workflows have then
emerged as a solution in which simulation and data analytics are intertwined
through shared computing resources, thus lower latencies.

Determining the best {\it allocation}, \ie how many resources to allocate to
each component of an \is workflow; and {\it mapping}, \ie where and at which
frequency to run the data analytics component, is a complex task whose
performance assessment is crucial to the efficient execution of \is
workflows. However, such a performance evaluation of different allocation and
mapping strategies usually relies either on directly running them on the
targeted execution environments, which can rapidly become extremely time- and
resource-consuming, or on resorting to the simulation of simplified models of
the components of an \is workflow, which can lack of realism. In both cases,
the validity of the performance evaluation is limited.

To address this issue, we introduce \simsitu, a framework for the faithful
simulation of \is workflows. This framework builds on the \simgrid toolkit and
benefits of several important features of this versatile simulation tool.  We
designed \simsitu to reflect the typical structure of \is workflows and thanks
to its modular design, \simsitu has the necessary flexibility to easily and
faithfully evaluate the behavior and performance of various allocation and
mapping strategies for \is workflows. We illustrate the simulation capabilities
of \simsitu on a Molecular Dynamics use case. We study the impact of different
allocation and mapping strategies on performance and show how users can
leverage \simsitu to determine interesting tradeoffs when designing their \is
workflow. \end{abstract}

\newpage
\pagenumbering{arabic}
\section{Introduction}

The tremendous volumes of data generated by numerical simulations in various
scientific domains such as molecular dynamics, climate modeling, biology, or
astrophysics, led to a fundamental redesign of application workflows. Due to
the growing discrepancy between storage subsystems performance and computing
power in extreme-scale supercomputers, moving large volumes of
data from computational resources to disks may have a dramatic impact on
performance~\cite{osti_1417653}.
This makes the classical post-hoc analysis of simulation outputs
highly inefficient. To overcome these issues, \is workflows intertwine
simulation and data analytics within shared computing resources. The objective
is to take advantage of data locality by consuming partial simulation results
to periodically run the data analysis. Then, the final outcome of the workflow
execution will only be the result of the data analytics phase, which is usually
much smaller than the entire simulation data.

A common challenge to the many software frameworks that have been developed to
efficiently process \is workflows is to decide what is the best {\it
allocation}, \ie how many resources to allocate to each component of an \is
workflow, and {\it mapping}---where and at which frequency run the data
analytics component. This is a complex task whose performance assessment is
crucial to the efficient execution of \is workflows. However, such a
performance evaluation of different allocation and mapping strategies usually
relies either on directly running them on the targeted execution environments
or on resorting to the simulation of simplistic models of the components of an
\is workflow. The former requires access to clusters and an important time to
tune the framework with regard to the hardware and software available on the
target machine and thus can rapidly become extremely time- and
resource-consuming, while the latter can lack of realism. In both cases, the
validity of the performance evaluation is usually limited to a narrow set of
configurations.

In this paper, we present \textbf{\simsitu}, a framework for the faithful
simulation of \is workflows. \simsitu builds on the popular \simgrid
toolkit~\cite{simgrid} and benefits of several key features of this versatile
simulation framework. Indeed, \simgrid enables the simulation of large-scale
distributed applications in a way that is accurate (via validated simulation
models), scalable (ability to run large scale simulations on a single computer
with low compute, memory, and energy footprints), and expressive (ability to
simulate arbitrary platform, application, and execution scenarios). We designed
\simsitu to reflect the typical structure of \is workflows with three distinct
modules that respectively, (i)~simulate the unmodified simulation component of
an \is workflow; (ii)~mimic the behavior of an underlying Data Transport Layer
(DTL); and (iii)~abstract the data analytics processing.

Thanks to this modular design, \simsitu has the necessary flexibility to easily
and faithfully evaluate the behavior and performance of various allocation
policies for \is workflows. We illustrate the simulation capabilities of
\simsitu on a Molecular Dynamics  use case. We study the impact of different
allocation and mapping strategies on performance and show how users can
leverage \simsitu to determine interesting tradeoffs when designing their \is
workflow.

The remaining of this paper is organized as follows. Section~\ref{sec:related}
presents the related work. In Section~\ref{sec:archi}, we describe the
architecture of the \simsitu framework and detail its different features and
advantages.  In Section~\ref{sec:use-case}, we describe how we used \simsitu to
simulate a Molecular Dynamics (MD) \is workflow. Thanks to this implementation,
we evaluate different \is allocation and mapping strategies in
Section~\ref{sec:eval}. Finally, Section~\ref{sec:ccl} summarizes our
contributions and presents some future research directions.
 \section{Related Work}
\label{sec:related}

The evaluation of the performance of the execution of \is workflows, and in
particular that of different allocation and mapping strategies for both the
simulation and data analytics components, is a complex and multi-parametric
problem.  Different approaches have been proposed in the literature to
ascertain the performance gains brought by an \is (or \intransit) execution of
a given scientific workflow application and determine the best configuration
deployment of its components on a given target platform. We distinguish these
approaches depending on whether they rely on actual experiments~\cite{flexio,
  Malakar15, Malakar16, Sun15, Subedi19} or resort to
simulation~\cite{Lorhmann17,gaupy19,Do21} to evaluate the performance of \is
workflows. The former is intrinsically time- and resource-consuming while the
latter may suffer from simplification biases when the abstract versions of the
\is workflow components are developed.

To limit the cost of the experiments required to perform performance
evaluations, some works focused on a particular component of the \is 
workflow--the Data Transport Layer (DTL) that connects the simulation and the
analysis.  In such cases, series of experiments are conducted either using the
real application~\cite{flexio} or by leveraging data access traces to mimic an
application behavior~\cite{Sun15, Subedi19}.  Another approach consisted in
reducing the experimental configuration space to a selected set of promising
configurations. For instance, Malakar~{\it et al.} tackled the allocation and
mapping of an \is workflow as an optimization problem expressed as a
Mixed-Integer Linear Program~\cite{Malakar15,Malakar16}. Solving this
optimization problem results on a set of feasible \is analyses whose
performance has been assessed through experimentation in an actual platform.
Lorhmann~{\it et al.} leveraged surrogate models (\ie proxy-applications) of
expensive numerical simulation codes~\cite{Lorhmann17} to abstract application
models. While this approach substantially reduces the cost of the experiments,
it still captures the most important features of the considered application.

Only a few recent attempts has leveraged simulation to enable the exploration
of the \is parameter space.  Aupy~{\it et al.} designed a numerical
simulator~\cite{gaupy19} that measures evaluation metrics for scheduling
decisions by solving optimization problems on resource allocation and
partitioning for an \is analysis set. The simulator used a predetermined set of
simulation parameters to study the impact of the \is analyses that are
scheduled on the performance of the entire \is execution.  In a previous 
work~\cite{Do21}, we
extrapolated benchmarking performance of realistic MD simulation engines to
create a synthetic MD simulation that emulates the behavior of MD
simulations while using fewer computing resources. Our proposed synthetic MD
simulation considered computation units as timing delays without actually
performing any computing operations.

To the best of our knowledge, this is the first work that proposes a faithful
and scalable simulation framework that models the behavior of all the
components of \is workflows.
 \section{\simsitu Architecture}
\label{sec:archi}

\begin{figure}[hbtp]
  \centering
  \includegraphics[width=\linewidth]{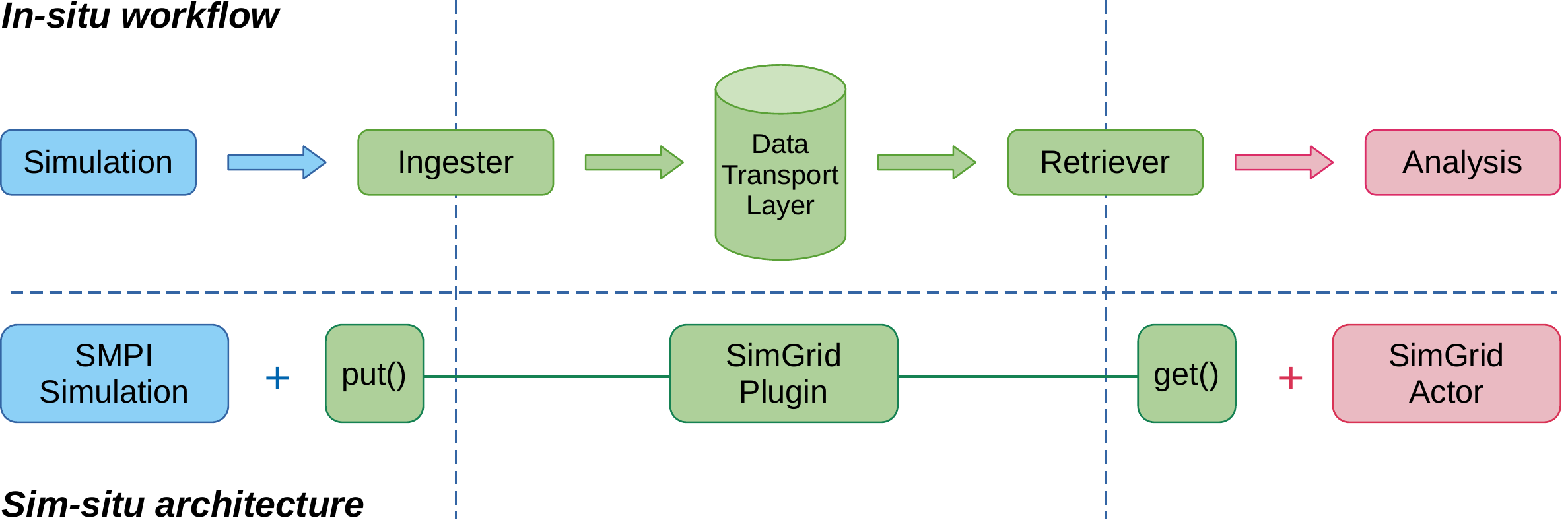}
  \caption{Software architectures of a generic \emph{in-situ} workflow (top)
    and the \simsitu framework (bottom).}
  \label{fig:archi}
\end{figure}

In this paper, we focus on a particular type of \is workflows whose structure
is depicted in the top part of Figure~\ref{fig:archi}. These workflows feature
the following main components:
\begin{itemize}
\item A \emph{Simulation} component that performs domain-specific computations,
  generally in an iterative process. This component periodically, \ie after a
  predefined number of iterations, produces some scientific data;
\item A \emph{Data Analytics} component that consumes the data generated by the
  simulation component and applies one or several analysis kernels;
\item A \emph{Data Transport Layer}~(DTL), whose complexity depends on the
  degree of coupling between the two former components and the resources they
  are allocated to, is responsible for efficient data transfers.
\end{itemize}

To faithfully simulate such \is workflows, we introduce the \simsitu framework
that leverages several features of the \simgrid
toolkit~\cite{simgrid}. \simgrid is an open-source versatile framework for
developing simulators of distributed applications that are executed on
heterogeneous distributed platform. One of the key strengths of \simgrid is to
not trade accuracy for scalability. Its fast simulation models have been
theoretically and experimentally evaluated and validated~\cite{simgrid-tomacs}
and make it possible to run large-scale simulations on a single machine.  The
way we simulate the components of an \is workflow is illustrated by the bottom
part of Figure~\ref{fig:archi} and detailed hereafter.

\pp{Simulation Component} To maximize the realism of the simulation of a
parallel application, such as the simulation component on an \is workflow, an
appealing approach is to directly simulate the unmodified code of the parallel
application itself. This ensures that the simulation not only captures the
computations executed on the different processes but also the exact
communication pattern of the application.

This approach is at the origin of SMPI, a flexible simulator of MPI
applications~\cite{smpi} that comes with the \simgrid distribution. SMPI works
seamlessly with unmodified MPI programs written in C, C++, or FORTRAN. As part
of its integration testing
effort\footnote{\url{https://framagit.org/simgrid/SMPI-proxy-apps}} SMPI
simulates the execution of multiple proxy- and full-scale applications and HPC
runtimes.  The code is compiled using one of the language specific compilers
shipped with \simgrid. For instance, assuming the program is written in C, one
simply compiles it with \smpicc, instead of \mpicc, and executes it with
\smpirun, instead of \mpirun.  The only difference is that \smpirun takes one
extra command-line argument, \texttt{-platform}, which allows the user to
describe the hardware platform on which the execution of the MPI program is to
be simulated. This platform description specifies a network topology between
compute nodes, where network links have specified latencies and bandwidths, and
compute nodes have specified compute speeds, and numbers of cores.

The basic principle behind SMPI simulations is as follows.  The code of the MPI
program is executed as is, but the MPI ranks actually execute as threads in a
single process, and thus share the same address space. Each time an MPI
function is called, control is handed off to SMPI where network operations are
replaced by simulated delays. These delays are computed using the simulation
models at the core of \simgrid~\cite{smpi-pmbs}.  Each block of code in between
two MPI calls is benchmarked on the machine used to execute the
simulation. This is possible because, with SMPI, MPI ranks execute as threads
in mutual exclusion.  These benchmarked execution times can then be scaled and
simulated as compute delays that correspond to the compute speeds of the nodes
in the simulated platform. In this way, SMPI simulates both communication and
computation operations as computed delays.

\pp{Data Analytics} The data analytics component of an \is workflow is usually
specific to a given scientific study. Moreover, its complexity can vary greatly
depending on what knowledge scientists want to extract from the simulation
data. It can range from a simple computation of some performance metrics to
help steer the simulation to a complex visualization of the current state of
the simulation. Then, to simulate this component in \simsitu we decide to
simply abstract its execution time within one or several \simgrid
\emph{actors}. A \simgrid actor corresponds to a simulated process that can
consume some simulated \emph{resources} (\eg CPU time, network bandwidth, or
storage space) by performing some simulated \emph{activities} (\eg executing a
computation, doing some I/O, or communicating with another actor).

Isolating and abstracting the data analytics component within actors out of the
MPI world offers a great flexibility to \simsitu and let us envisage multiple
scenarios. First, decoupling the simulation from the analysis in \simsitu's
code allows us to easily decide where to map these actors. Indeed, an actor can
be started on any node of the simulated cluster, that execution location being
specified at the creation of the actor, or predefined in the description of an
initial static deployment of the different actors. Second, abstracting the
analysis as an activity consuming CPU resources allows us to easily study
various simulation over analysis time ratios and determine what should be the
best allocation and mapping strategy in each scenario. Last, it is also
possible to spawn new actors, stop existing ones, or even migrate them from one
node to another while the \simgrid simulation is running. This would allow
\simsitu to study complex scenarios where the analysis load evolves along time.

\pp{Data Transport Layer} To simulate data exchanges between the simulation and
the data analytics components, we opted for the development of a \simgrid
plugin. A \simgrid plugin is a standalone piece of code that composes some of
basic concepts exposed by the \simgrid API to offer a higher level of
abstraction useful in a specific context. In this paper, the proposed plugin
relies on data structures accessed through a Producer-Consumer synchronization
mechanism commonly used in actual DTLs\cite{dataspaces, dimes}. Using this
plugin only requires to include an extra header file in the application code to
use the functions it provides.

Two features of this plugin are especially interesting in the context of
\simsitu. First, it offers two different internal implementations of the
message queue that allow us to consider three different communication modes
between the main components of an \is workflow. The former is a standard
\emph{queue} (of configurable size). In this case, data exchanges are
instantaneous, \ie do not induce any advance of the simulated clock, but
respect the flow dependencies between the producers and consumers. This allows
us to study scenarios where the analysis component has a direct and seamless
access to the data produced by the simulation component, or to only focus on
the computational element of the \is workflow and study the impact of different
allocation and mapping schemes on that element alone.

The latter implementation leverages the concept of \emph{mailbox} used by
\simgrid to implement inter-process communications. It acts as a rendez-vous
point between a sender and a receiver processes. When both meet on that
rendez-vous point, the actual communication starts. \simgrid mailboxes use a
queue to store unmatched communications, \ie when one side is waiting for the
other, which ensures the respect of flow dependencies. An interesting feature
of mailboxes is that depending on where the processes are located, we can
simulate two different ways to exchange data. If the producer and the consumer
are located on the same compute node, and thus sharing a memory space, the
simulated communication will go through the loopback of this node. This allows
us to simulate a memory copy of data between the simulation and data analytics
components. Conversely, if the producer and the consumer are located on
different nodes, the communication will go over the interconnection
network. This makes it possible to easily compare \is and \intransit scenarios.

The second interesting feature of the proposed plugin is that accesses to the
simulated DTL can be done either in synchronous or asynchronous modes. Again,
this offers more flexibility and broaden the range of scenarios that can be
studied with \simsitu.
 \section{Application to MD In-situ Workflows}
\label{sec:use-case}

To assess the accuracy and illustrate the flexibility of the proposed \simsitu
framework, we consider the simulation of a Molecular Dynamics (MD) \is
workflow. MD aims at studying the evolution of molecular systems at the atomic
scale and, is one of the most prominent types of numerical simulations
currently running on extreme-scale systems.

\pp{Application} More precisely, we relied for our experiments on the \emd
proxy-application~\cite{ExaMiniMD, ExaMiniMD-github} which is part of the
Exascale Computing Project Proxy App Suite v4.0~\cite{ECP-proxy-apps}.  \emd
captures both the computation and communication schemes that are implemented in
the classical MD code LAMMPS ~\cite{LAMMPS}. As other proxy-applications, \emd
shows a good balance between having a compact and manageable code and
representing the main performance concerns of MD applications.  \emd belongs to
the family of the all-atom MD simulations. It computes floating-point intensive
pairwise atom-atom unbounded interactions over a certain period of time. The
simulated system corresponds to a set of particles distributed in a 3D
volume. The main loop computes the trajectories of the particles according to a
Verlet time integration method and the short-range forces between particles as
a Lennard-Jones potential. The parallelization of this MD problem follows a
typical domain decomposition approach. Each MPI rank manages a sub-volume and a
halo to exchange with its neighbors periodically. The periodicity of these
exchanges can be set to {\tt X} by adding \verb+neigh_modify every X+ to the
input file. All the data exchanges in the simulation component rely on
point-to-point MPI communications with asynchronous receives.

The data analytics component of \emd consists in periodically computing the
temperature, potential energy, and kinetic energy of the system. It is entangled
with the simulation code and uses the same process mapping. Each MPI rank
locally computes the different metrics and then enters a global
\verb+MPI_Allreduce+ function that leads to the final value. The periodicity of
this analysis is set by the {\tt thermo} input parameter. If no value is given,
the computation of these values are only done before and after the main
simulation loop.

\pp{Experimental Platform} Our set of experiments and simulations were
conducted on the {\it dahu} cluster of the Grid'5000 experimental testbed. This
cluster consists of 32 nodes that comprise two Intel Xeon Gold 6130 CPUs with
16 cores each and 192 GiB of memory. These nodes are interconnected through a
10 Gb/s Ethernet network.  We leverage an existing thorough calibration of the
SMPI network model for this same cluster~\cite{these-tom} to ensure our
simulated results are accurate. This calibration runs a series of
tests on a limited number of nodes to assess the performance of point-to-point
communications and saturate a switch. It can then be used to
extrapolate the size of a given cluster beyond its actual number of nodes.

We used the git version of \emd, compiled with g++ v8.3.0 and linked to Kokkos
v3.3.01 and OpenMPI v3.1.3. We used the {\tt Serial} Kokkos device with one MPI
rank per core as we observed a performance degradation with the {\tt Pthread}
and {\tt OpenMP} devices combined with MPI.  For the simulation of \emd, we
relied on \simgrid v3.29.

\subsection{Simulation Component}
\label{sec:XP-sim}
In this section, we analyze the performance and accuracy of the simulation with
SMPI of the unmodified code of the \emd application. Enabling this simulation
only requires minimal modification to the {\tt Makefile} file to indicate that
the compiler to use is {\tt smpicxx}.

An additional and optional modification of the code of \emd can be made to
drastically reduce the simulation time. SMPI offers to replace
time-consuming computational parts of the simulated application by delays to
speedup the simulation. These delays are estimated by sampling the execution
time of a given kernel or loop body either for a predefined number of times or
until the standard deviation of the samples is under a given threshold. All the
subsequent calls are then replaced by the average execution time of these
samples. Finally, this sampling can be done either at a {\it local} scale, \ie
each MPI rank determines its own delay from the samples it executed, or at a
{\it global} scale, \ie the delay is determined from samples executed by all
the MPI ranks.

We used this feature of SMPI on the most time-consuming kernel of \emd which is
the call to \verb+ForceLJNeigh::compute+. This particular compute-bound kernel
represents 69\% of the execution time of the
application~\cite{ECP-proxy-apps-FY18}. This corresponds to a 1-line
modification of the code to call the SMPI sampling macro with its
parameters. We chose to run 150 samples with a standard deviation threshold of
0.002 in our experiments.

\begin{figure}[hb!]
  \centering
  \includegraphics[width=.94\linewidth]{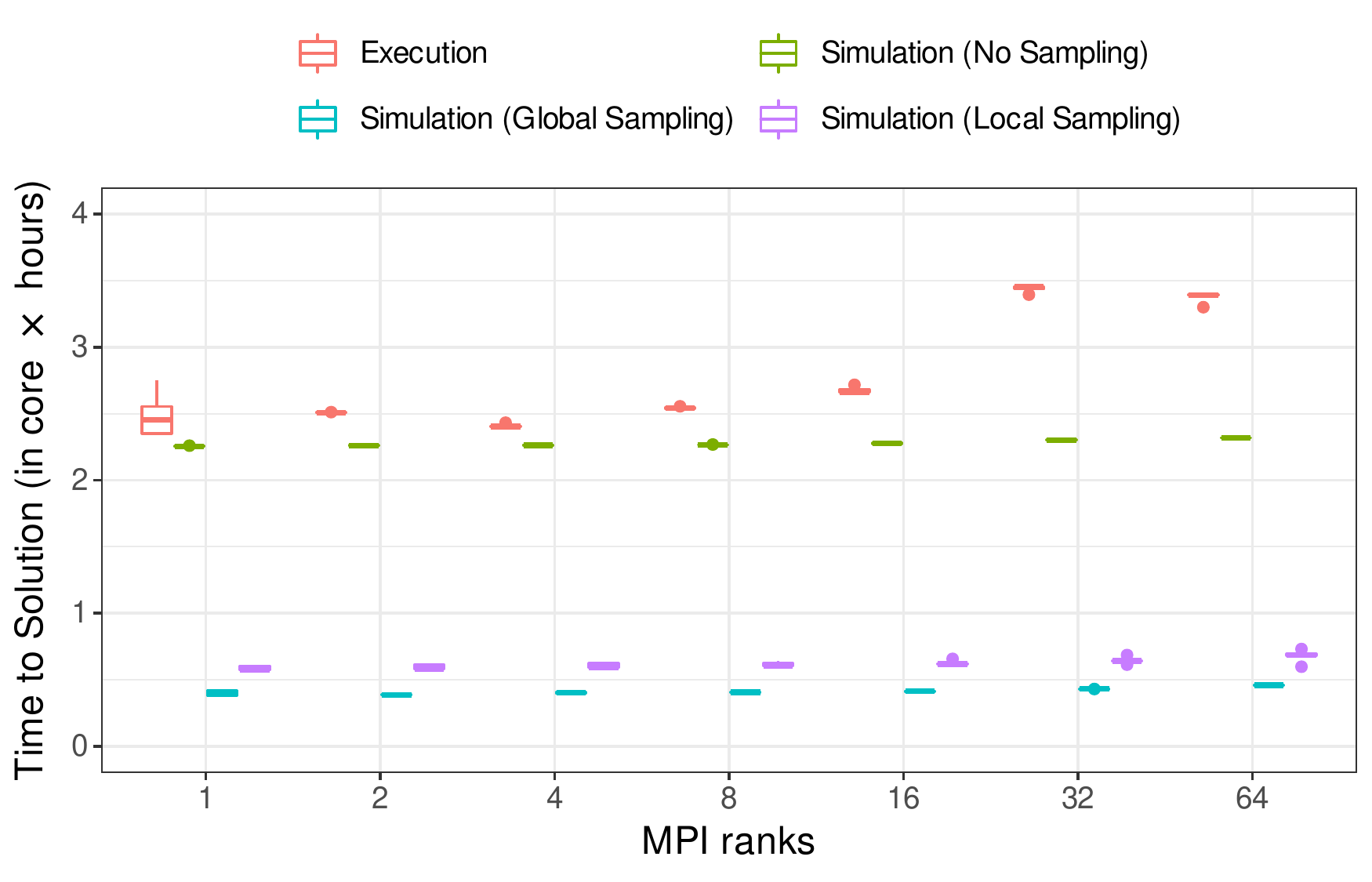}
  \caption{Time to run or simulate (with or without
    kernel sampling) the execution of a 3D Lennard-Jones melt on a
    $70^{3}$ region with \emd. Each rank runs a single Kokkos
    thread and is mapped on a different core.}
  \label{fig:time_to_solution}
\end{figure}

\pp{Characterization}
Figure~\ref{fig:time_to_solution} compares the time needed to run 8,000
iterations of a 3D Lennard-Jones melt on a $70\times70\times70$ region with
\emd and the time needed to simulate this execution with SMPI for different
numbers of MPI ranks. Data exchanges among ranks occur every 20 iterations and
the analytics phase is triggered every 50 iterations. We map~an~MPI rank per
core and use a single Kokkos thread per MPI rank.

We can see that the number of core $\times$ hours needed to solve this problem
instance remains stable as the number of MPI ranks increases, \ie the actual
execution completes faster with higher rank counts. The simulation of \emd with
SMPI runs on a single core and takes the same time to complete whatever the
number of MPI ranks used. This simulation time is commensurate to that of the
actual execution, but uses much less computing resources.  Activating the
kernel sampling, either local or global, in the simulation reduces the time to
solution by a factor of 5, thus results can be obtained in about 25-30 minutes
instead of 2.5 hours.

\begin{figure}[hbtp]
  \centering
  \includegraphics[width=.94\linewidth]{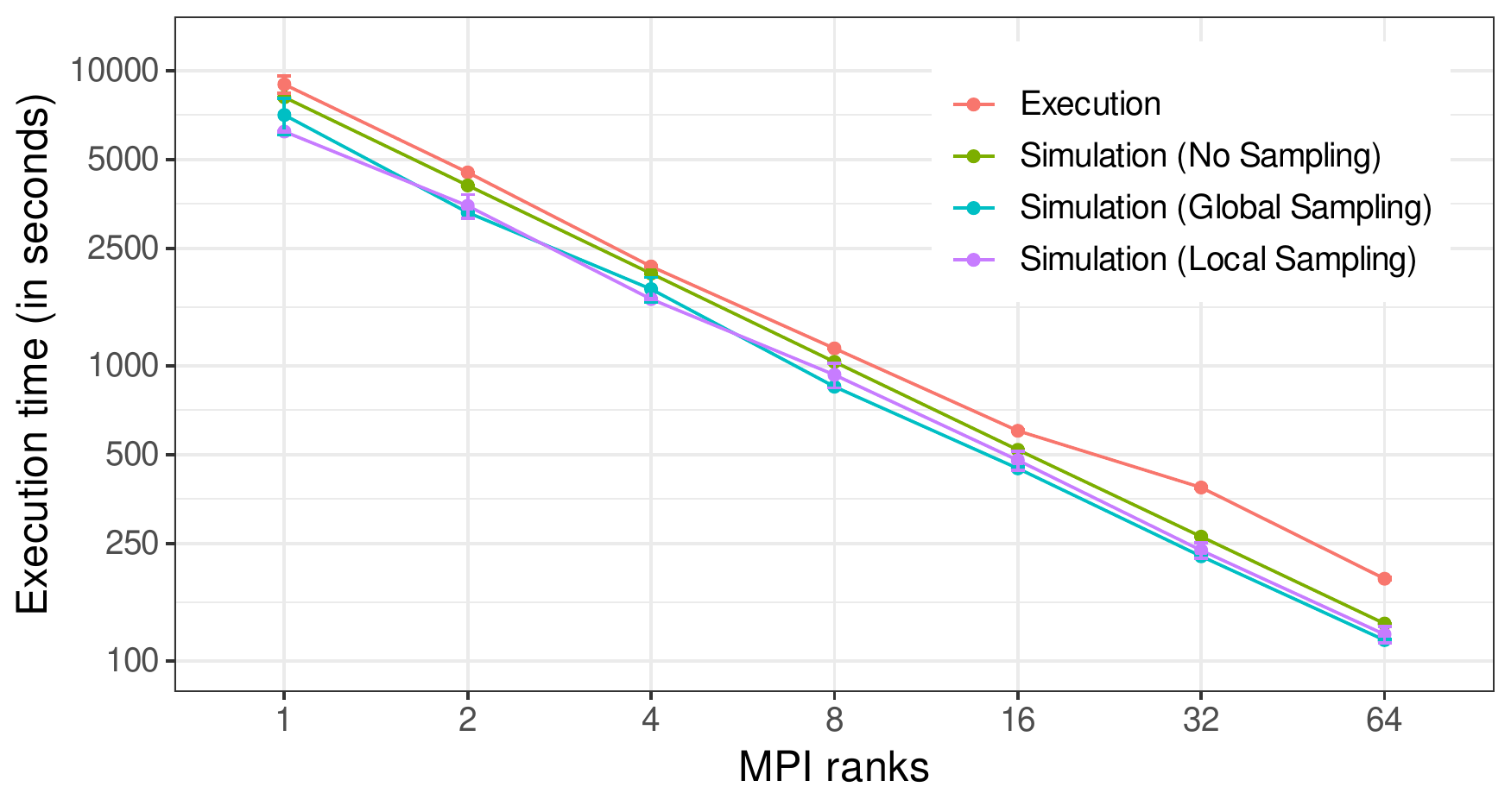}
  \caption{Accuracy of the simulation of 3D Lennard-Jones melt
    on a $70^{3}$ region (with or without kernel sampling) when varying
    the number of MPI ranks.}
  \label{fig:accuracy_serial}
\end{figure}

\pp{Accuracy}
Figure~\ref{fig:accuracy_serial} shows the accuracy of the simulation of an
unmodified version of \emd with SMPI and the impact of kernel sampling on this
accuracy. SMPI is able to correctly reflect the behavior of the simulated
application with less variability and a reasonable error. Activating the kernel
sampling, be it local or global, slightly degrades the accuracy of the
simulation. However, this degradation remains stable as the number of ranks
increases. Then, it can easily be taken into account when assessing the
performance of a given \is scheduling strategy with \simsitu.

\begin{figure}[hbtp]
  \centering
  \includegraphics[width=.94\linewidth]{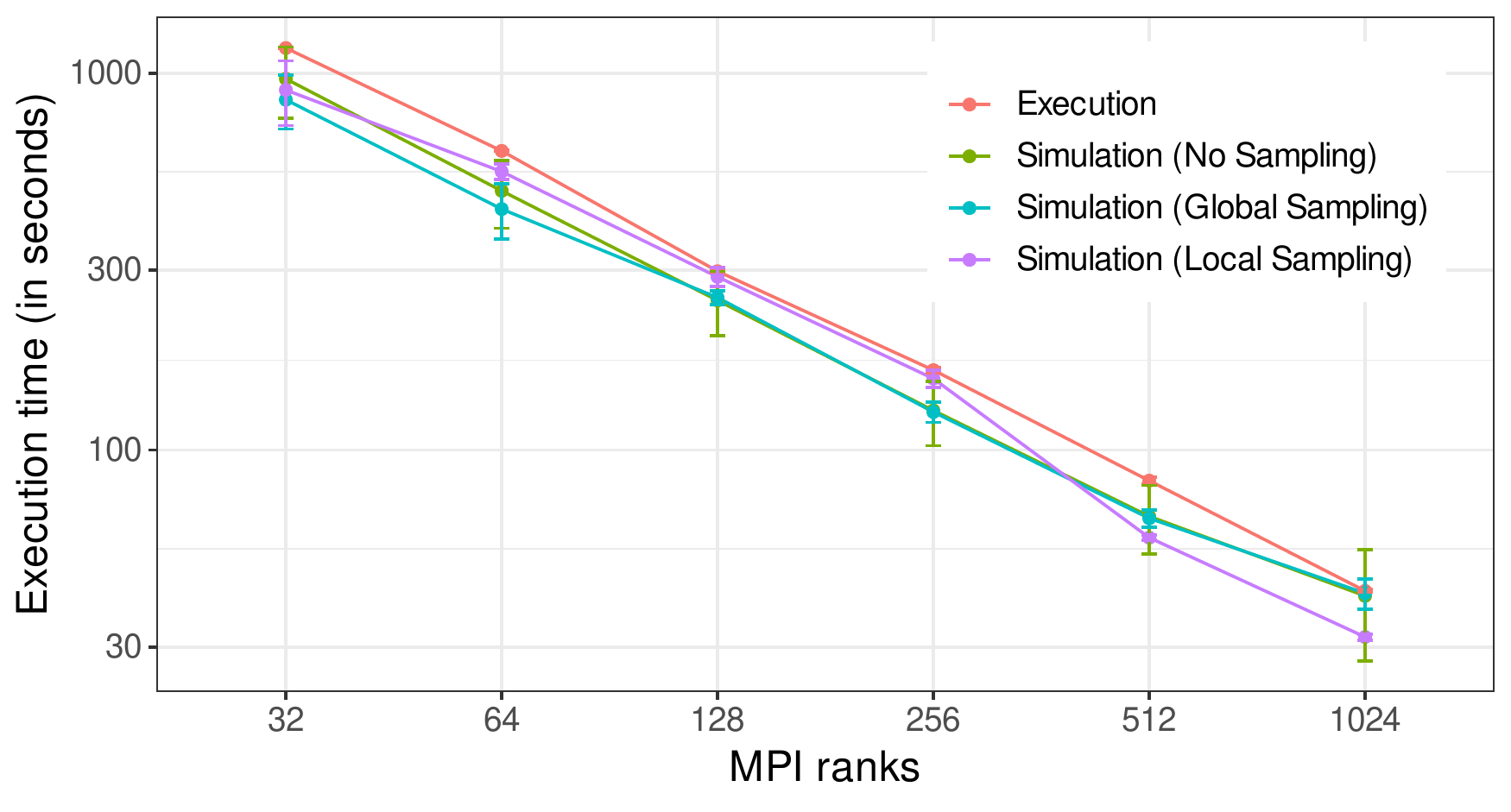}
  \caption{Accuracy of the simulation of 3D Lennard-Jones melt
    on a $90^{3}$ region (with or without kernel sampling) when varying
    the number of MPI ranks.}
  \label{fig:accuracy_serial_large}
\end{figure}

To assess the quality of our simulation on larger core counts, scaling up to
the full size of the target platform (\ie 32  nodes and 1,024 cores), we
simulated a larger problem instance with a $90 \times 90 \times 90$ region and
$12,000$ iterations. Figure~\ref{fig:accuracy_serial_large} shows that the
accuracy of the local sampling version drops from 512 cores while the global
sampling version and the simulation without sampling remain accurate.

\subsection{Data Analytics Component}
\label{sec:XP-ana}
Data analytics is the most versatile component of \is workflows. Users can act
on many parameters related to this component when designing and executing their
workflows. Depending on what they want to observe or steer during the execution
of the simulation component, the compute cost and complexity of the data
analytics function, the volume of data to transfer from simulation to
analytics, and the frequency of the analysis can change from one run to
another.  Combining these variable parameters leads to interesting questions
such as ``Is it more efficient to run frequent but light analyses or scarce but
heavy ones?". Moreover, opting for a given configuration will have a direct
impact on performance and may benefit of a particular allocation and mapping
scheme.

\pp{Configuration} To reflect this versality of the data analytics component
and help users in determining the best allocation and mapping for a given
configuration, we decided to augment \emd with an extra command line flag
(\verb+--analysis+). This allows us to configure the \is version of \emd
without having to recompile the application.  This flag requires six
parameters:
\begin{itemize}
\item The number of analytics actors to spawn;
\item A file, similar the MPI hostfile, that describes the mapping of the
  analytics actors;
\item The cost per particle of the analytics. The analytics actors then
  multiply this value by the number of analyzed particles to determine the
  amount of work they have to simulate;
\item A computing scaling factor. This parameter allows us to
  artificially increase the analytics cost to study different what-if
  scenarios;
\item The size per particle of the data transferred from the simulation
  component to the analytics component. Similarly, this value is multiplied by
  the number of analyzed particles;
\item A data transfer scaling factor. This parameter allows us to
  artificially increase the transferred data size to study different what-if
  scenarios.
\end{itemize}

\pp{In-situ Analytics} We slightly modified the code of \emd to plug the
proposed simulated analytics component. It is implemented as an external shared library
that is linked to \emd at compile time and is made of two different types of
\simgrid actors.  The behavior of the {\it analytics} actors, whose number can
be configured on command line, is described by
Algorithm~\ref{alg:analysis}. They are in charge of the simulated execution of
the analytics function with is computation of the different metrics computed
every {\tt thermo} steps.

\begin{algorithm}[hbtp]
  \caption{Analytics actor}
  \label{alg:analysis}
  \begin{algorithmic}[1]
    \LOOP
    \STATE Get system state from the DTL
    \IF{Poisoned value}
    \IF{Last actor running}
    \STATE Send poisoned value to metric collector
    \ENDIF
    \STATE Return
    \ENDIF
    \STATE Compute analytics 
    \STATE Send computed metrics to metric collector
    \ENDLOOP
  \end{algorithmic}
\end{algorithm}

Each {\it analytics} actor runs an infinite loop in which it waits for data to
analyze to be available in the DTL. When it is the case, the actor simulates
the data analytics function. Thanks to the flexible and
expressive API of SimGrid, simple functions that simulate the execution
of the amount of work given as parameter at the compute speed of the node the
actor is mapped on, or more complex data analytics, including complex
communication patterns and multi-node allocations, can be simulated. In the
latter case, communications of the simulation and analytics components share
the same network and contention will be captured by \simsitu if some occurs.
We simulate the three functions of \emd analytics component using the former 
scenario.

Then the analytics actor asynchronously sends dummy results to the {\it metric
  collector} actor, and waits again for new data. At the end of the execution
of the simulation component, a poisoned value is sent to all the analytics
actors to properly stop them. The last actor running then stops the metric
collector by sending it a poisoned value (lines 4-5).

The objective of the {\it metric collector} (Algorithm~\ref{alg:collector}) is
to simulate the accumulation of the metrics done in the analytics phase. As the
analytics is decoupled from the MPI simulation in \simsitu, a communication
scheme, that only implies the \simgrid actors and not the MPI ranks, can be
implemented.

\begin{algorithm}
  \caption{Metric collector actor}
  \label{alg:collector}
  \begin{algorithmic}[1]
    \LOOP
    \STATE n\_collected\_values = 0
    \REPEAT
    \STATE Get metrics from analytics actors
    \IF{Poisoned value}
    \STATE Return
    \ENDIF
    \STATE Accumulate metrics
    \UNTIL{n\_collected\_values = n\_ranks}
    \FORALL{n\_ranks}
    \STATE Put a copy of accumulated metrics into the DTL
    \ENDFOR
    \ENDLOOP
  \end{algorithmic}
\end{algorithm}

This actor simply waits for having received as many individually computed
metrics to accumulate as there are ranks executing the simulation
component~(lines 3-8). As the number of analytics actors can be smaller than
the number of MPI ranks, an actor can send more than one set of metrics to the
metric collector. Once all the metric values for a given analytics phase have
been collected, this actor puts as many copies of the accumulated values into
the DTL~(lines 10-11), so that each MPI rank can retrieve one set of metrics
and pursue its execution of the simulation component.

\subsection{Data Transport Layer}
\label{sec:DTL}

\begin{figure}[hbtp]
  \centering
  \includegraphics[width=.665\linewidth]{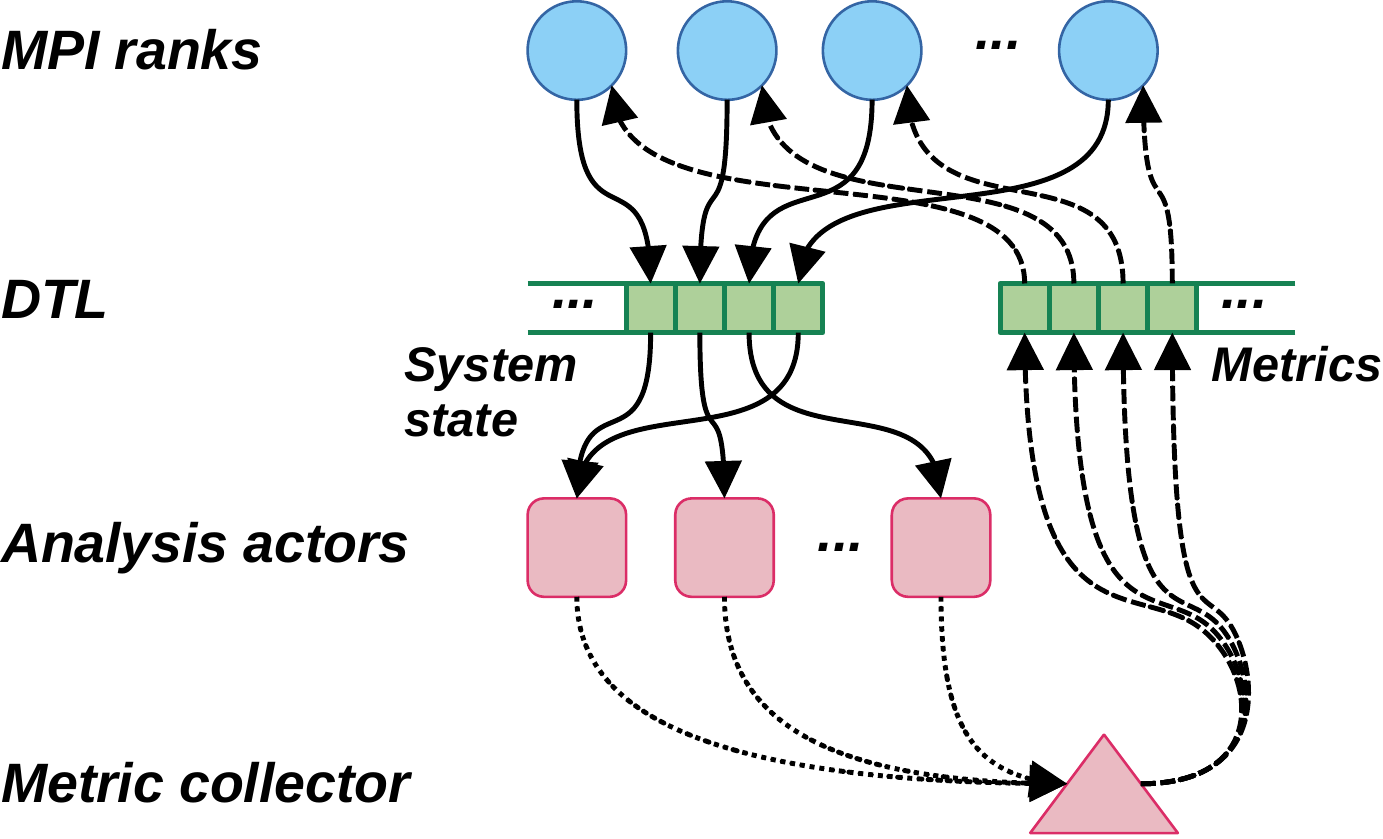}
  \caption{Data exchanges between the Simulation and Analytics components
    through the Data Transport Layer.}
  \label{fig:DTL-exchanges}
\end{figure}

The structure of the proposed Data Transport Layer~(DTL) and the data exchanges
between the simulation and analytics components of the \is workflow are
illustrated by Figure~\ref{fig:DTL-exchanges}. It is organized around two
distinct message queues. The former stores the current system states sent by
each of the MPI ranks to the analytics actors (plain arrows) while the latter
stores the metrics computed by the analytics actors and aggregated by the metric
collector that are sent back to the MPI ranks (dashed arrows). The
communications between the analytics and metric collector actors rely on a
standard \simgrid mailbox (dotted arrows), and are thus outside the MPI world.

These two message queues are accessed through a Producer-Consumer mechanism
provided by \simgrid. This reduces the amount of synchronization needed between
the simulation and the analytics components to a minimum. The MPI ranks ingest
their current system state into the DTL in a fire-and-forget mode and then
immediately proceed with the next iteration of the main MD simulation
loop. They will then block to retrieve the analysis results, but only after
{\tt thermo} iterations, \ie before having to start a new analysis.  This
Producer-Consumer mechanism also improves the flexibility of \simsitu as it
allows us to start any number of analytics actors without having to further
modify the original code of the application.

 \section{Evaluation of In-Situ Allocation and Mapping Strategies with \simsitu}
\label{sec:eval}

In this section, we introduce a model and a metric to assess the performance
of different allocation and mapping schemes and illustrate how the \simsitu
framework can be used\footnote{A reproducibility artifact for this paper is
  available {\href{https://figshare.com/s/e71c57503025e2389779}{\underline{online}}.}}
to study their impact on the performance of the MD \is
workflow detailed in the previous section.

\subsection{Modeling the \emd In-Situ Workflow}

The following execution model is adapted from the one proposed
in~\cite{Do21}. The main difference lies in the data transfer pattern between
the simulation and analytics components. In the model in~\cite{Do21}, the
simulation component only sends data to the analytics once, with some buffer
constraints imposed by the DTL, while in the considered \is workflow, the the
results of analytics component have to be received before starting a new
analysis.

Such a model can help users to obtain a first approximation of the impact of
the different configuration parameters of the \is workflow (\ie cost and
frequency of analysis) and of the allocation and mapping scheme (\ie number of
nodes and core allocation ratio) on the performance of the application. Then,
faithful simulations with \simsitu can complement this approximation by taking
into account more complex phenomena that are captured by the simulator (\eg
contention over the network, load imbalance, etc.).

\begin{figure}[htbp]
\subfloat[\idleS Scenario: $\get{i} + \ana{i} >\simu{i}+\ingest {i}$]{
\centering
\resizebox{\linewidth}{!}{
   \begin{tikzpicture}
\newcommand{\widthSim}{1.1}
\newcommand{\widthIng}{0.35}
\newcommand{\widthRead}{0.35}

\newcommand{\widthAna}{1.3}
\newcommand{\widthSend}{0.24}
\newcommand{\widthCollect}{0.24}

\newcommand{\height}{0.5}
\newcommand{\overhead}{0.5}

\newcommand{\widthDTLSysS}{0.5}
\newcommand{\widthDTLMetrics}{0.3}
\newcommand{\widthIdleA}{0.5}
\newcommand{\widthIdleS}{0.28}

\newcommand{\colorSim}{[fill=cyan!50]}
\newcommand{\colorAna}{[fill =red!50]}
\newcommand{\colorTransfer}{[fill=green!40]}
\newcommand{\colorIdleTime}{[fill=white]}

\draw\colorSim (0,\height) rectangle (\widthSim,2*\height) node[pos=.5, font=\fontsize{07}{0}\selectfont] {\simu{1}};
\draw\colorTransfer  (\widthSim,\height) rectangle (\widthSim+\widthIng,2*\height) node[pos=.45, rotate=-90,font=\fontsize{05.5}{0}\selectfont] {\ingest{1}};

\draw\colorTransfer  (\widthSim+\widthIng, 0) rectangle (\widthSim+\widthIng+\widthRead,\height) node[pos=.5, font=\fontsize{07}{0}\selectfont] {\get{1}};
\draw\colorAna (\widthSim+\widthIng+\widthRead,0) rectangle (\widthSim+\widthIng+\widthRead+\widthAna,\height) node[pos=.5, font=\fontsize{07}{0}\selectfont] {\ana{1}};

 \draw\colorTransfer  (\widthSim+\widthIng+\widthRead+\widthAna,0) rectangle (\widthSim+\widthIng+\widthRead+\widthAna+\widthSend,\height) node[pos=.52,rotate=-90, font=\fontsize{05}{0}\selectfont] {\send{1}};
 \draw\colorTransfer  (\widthSim+\widthIng+\widthRead+\widthAna+\widthSend,\height) rectangle (\widthSim+\widthIng+\widthRead+\widthAna+\widthSend+\widthCollect,2*\height) node[pos=.52, rotate=-90,font=\fontsize{05}{0}\selectfont] {\collect{1}};

\draw [dashed] (\widthSim+\widthIng+\widthRead+\widthAna+\widthSend,-\overhead) rectangle (\widthSim+\widthIng+\widthRead+\widthAna+\widthSend+\widthCollect,2*\height+\overhead);

\draw\colorSim (\widthSim+\widthIng,\height) rectangle (2*\widthSim+\widthIng,2*\height) node[pos=.5, font=\fontsize{07}{0}\selectfont] {\simu{2}};

\node[red] (IS1) at (2*\widthSim+\widthIng+\widthIdleS,1.5*\height) {\idle{2}{S}};

\draw\colorTransfer  (\widthSim+\widthIng+\widthRead+\widthAna+\widthSend+\widthCollect,\height) rectangle (\widthSim+2*\widthIng+\widthRead+\widthAna+\widthSend+
\widthCollect,2*\height) node[pos=.45, rotate=-90, font=\fontsize{05.5}{0}\selectfont] {\ingest{2}};

 \draw\colorTransfer  (\widthSim+2*\widthIng+\widthRead+\widthAna+\widthSend+\widthCollect, 0) rectangle (\widthSim+2*\widthIng+2*\widthRead+\widthAna+\widthSend+
\widthCollect,\height) node[pos=.5, font=\fontsize{07}{0}\selectfont] {\get{2}};
 \draw\colorAna (\widthSim+2*\widthIng+2*\widthRead+\widthAna+\widthSend+\widthCollect,0) rectangle (\widthSim+2*\widthIng+2*\widthRead+2*\widthAna+\widthSend+\widthCollect,\height) node[pos=.5, font=\fontsize{07}{0}\selectfont] {\ana{2}};

\draw\colorTransfer  (\widthSim+2*\widthIng+2*\widthRead+2*\widthAna+\widthSend+\widthCollect,0) rectangle (\widthSim+2*\widthIng+2*\widthRead+2*\widthAna+2*\widthSend+\widthCollect,\height) node[pos=.52,rotate=-90, font=\fontsize{05}{0}\selectfont] {\send{2}};
\draw\colorTransfer  (\widthSim+2*\widthIng+2*\widthRead+2*\widthAna+2*\widthSend+\widthCollect,\height) rectangle (\widthSim+2*\widthIng+2*\widthRead+2*\widthAna+2*\widthSend+2*\widthCollect,2*\height) node[pos=.52, rotate=-90, font=\fontsize{05}{0}\selectfont] {\collect{2}};

\draw [dashed] (\widthSim+2*\widthIng+2*\widthRead+2*\widthAna+2*\widthSend+\widthCollect,-\overhead) rectangle (\widthSim+2*\widthIng+2*\widthRead+2*\widthAna+2*\widthSend+2*\widthCollect,2*\height+\overhead);

\draw\colorSim (\widthSim+2*\widthIng+\widthRead+\widthAna+\widthSend+\widthCollect,\height) rectangle (2*\widthSim+2*\widthIng+\widthRead+\widthAna+\widthSend+
\widthCollect,2*\height) node[pos=.5, font=\fontsize{07}{0}\selectfont] {\simu{3}};

\node[red] (IS3) at (2*\widthSim+2*\widthIng+\widthRead+\widthAna+\widthSend+\widthCollect+\widthIdleS,1.5*\height) {\idle{3}{S}};

\draw\colorTransfer  (\widthSim+2*\widthIng+2*\widthRead+2*\widthAna+2*\widthSend+2*\widthCollect,\height) rectangle (\widthSim+3*\widthIng+2*\widthRead+2*\widthAna+2*\widthSend+2*\widthCollect,2*\height) node[pos=.45, rotate=-90, font=\fontsize{05.5}{0}\selectfont] {\ingest{3}};

 \draw\colorTransfer  (\widthSim+3*\widthIng+2*\widthRead+2*\widthAna+2*\widthSend+2*\widthCollect, 0) rectangle (\widthSim+3*\widthIng+3*\widthRead+2*\widthAna+2*\widthSend+2*\widthCollect,\height) node[pos=.5, font=\fontsize{07}{0}\selectfont] {\get{3}};
 \draw\colorAna (\widthSim+3*\widthIng+3*\widthRead+2*\widthAna+2*\widthSend+2*\widthCollect,0) rectangle (\widthSim+3*\widthIng+3*\widthRead+3*\widthAna+2*\widthSend+2*\widthCollect,\height) node[pos=.5, font=\fontsize{07}{0}\selectfont] {\ana{3}};

\draw\colorTransfer  (\widthSim+3*\widthIng+3*\widthRead+3*\widthAna+2*\widthSend+2*\widthCollect,0) rectangle (\widthSim+3*\widthIng+3*\widthRead+3*\widthAna+3*\widthSend+2*\widthCollect,\height) node[pos=.52,rotate=-90, font=\fontsize{05}{0}\selectfont] {\send{3}};
\draw\colorTransfer  (\widthSim+3*\widthIng+3*\widthRead+3*\widthAna+3*\widthSend+2*\widthCollect,\height) rectangle (\widthSim+3*\widthIng+3*\widthRead+3*\widthAna+3*\widthSend+3*\widthCollect,2*\height) node[pos=.52, rotate=-90, font=\fontsize{05}{0}\selectfont] {\collect{3}};

\draw [dashed] (\widthSim+3*\widthIng+3*\widthRead+3*\widthAna+3*\widthSend+2*\widthCollect,-\overhead) rectangle (\widthSim+3*\widthIng+3*\widthRead+3*\widthAna+3*\widthSend+3*\widthCollect,2*\height+\overhead);

\draw\colorSim (\widthSim+3*\widthIng+2*\widthRead+2*\widthAna+2*\widthSend+2*\widthCollect,\height) rectangle (2*\widthSim+3*\widthIng+2*\widthRead+2*\widthAna+2*\widthSend+2*\widthCollect,2*\height) node[pos=.5, font=\fontsize{05}{0}\selectfont] {\simu{4}};

\node[red] (IS4) at (2*\widthSim+3*\widthIng+2*\widthRead+2*\widthAna+2*\widthSend+2*\widthCollect+\widthIdleS,1.5*\height) {\idle{4}{S}};

\end{tikzpicture} }
   \label{fig:idleS}
}

\subfloat[\idleA scenario: $\simu{i}+\ingest {i} > \get{i} + \ana{i}$]{
\centering
\resizebox{\linewidth}{!}{
   \begin{tikzpicture}
\newcommand{\widthSim}{1.7}
\newcommand{\widthIng}{0.35}
\newcommand{\widthRead}{0.325}

\newcommand{\widthAna}{0.65}
\newcommand{\widthSend}{0.225}
\newcommand{\widthCollect}{0.24}

\newcommand{\height}{0.5}
\newcommand{\overhead}{0.3}

\newcommand{\widthIdleA}{0.25}

\newcommand{\colorSim}{[fill=cyan!50]}
\newcommand{\colorAna}{[fill =red!50]}
\newcommand{\colorTransfer}{[fill=green!40]}

\draw\colorSim (0,\height) rectangle (\widthSim,2*\height) node[pos=.5, font=\fontsize{07}{0}\selectfont] {\simu{1}};
\draw\colorTransfer  (\widthSim,\height) rectangle (\widthSim+\widthIng,2*\height) node[pos=.45, rotate=-90,font=\fontsize{05.5}{0}\selectfont] {\ingest{1}};

\draw\colorTransfer  (\widthSim+\widthIng, 0) rectangle (\widthSim+\widthIng+\widthRead,\height) node[pos=.5, font=\fontsize{07}{0}\selectfont] {\get{1}};
\draw\colorAna (\widthSim+\widthIng+\widthRead,0) rectangle (\widthSim+\widthIng+\widthRead+\widthAna,\height) node[pos=.5, font=\fontsize{07}{0}\selectfont] {\ana{1}};
 \draw\colorTransfer  (\widthSim+\widthIng+\widthRead+\widthAna,0) rectangle (\widthSim+\widthIng+\widthRead+\widthAna+\widthSend,\height) node[pos=.52,rotate=-90, font=\fontsize{05}{0}\selectfont] {\send{1}};

 \draw\colorTransfer  (2*\widthSim+\widthIng,\height) rectangle (2*\widthSim+\widthIng+\widthCollect,2*\height) node[pos=.52, rotate=-90,font=\fontsize{05}{0}\selectfont] {\collect{1}};

\draw [dashed] (2*\widthSim+\widthIng,-\overhead) rectangle (2*\widthSim+\widthIng+\widthCollect,2*\height+\overhead);

\draw\colorSim (\widthSim+\widthIng,\height) rectangle (2*\widthSim+\widthIng,2*\height) node[pos=.5, font=\fontsize{07}{0}\selectfont] {\simu{2}};

 \node[red] (IA1) at (\widthSim+\widthIng+\widthRead+\widthAna+\widthSend+\widthIdleA,0.5*\height) {\idle{1}{A}};
 
 \draw\colorTransfer  (2*\widthSim+\widthIng+\widthCollect,\height) rectangle (2*\widthSim+2*\widthIng+\widthCollect,2*\height) node[pos=.45, rotate=-90, font=\fontsize{05.5}{0}\selectfont] {\ingest{2}};

\draw\colorTransfer  (2*\widthSim+2*\widthIng+\widthCollect, 0) rectangle (2*\widthSim+2*\widthIng+\widthCollect+\widthRead,\height) node[pos=.5, font=\fontsize{07}{0}\selectfont] {\get{2}};
\draw\colorAna (2*\widthSim+2*\widthIng+\widthCollect+\widthRead,0) rectangle (2*\widthSim+2*\widthIng+\widthCollect+\widthRead+\widthAna,\height) node[pos=.5, font=\fontsize{07}{0}\selectfont] {\ana{2}};
\draw\colorTransfer  (2*\widthSim+2*\widthIng+\widthCollect+\widthRead+\widthAna,0) rectangle (2*\widthSim+2*\widthIng+\widthCollect+\widthRead+\widthAna+\widthSend,\height) node[pos=.52,rotate=-90, font=\fontsize{05}{0}\selectfont] {\send{2}};

\draw\colorTransfer  (3*\widthSim+2*\widthIng+\widthCollect,\height) rectangle (3*\widthSim+2*\widthIng+2*\widthCollect,2*\height) node[pos=.52, rotate=-90, font=\fontsize{05}{0}\selectfont] {\collect{2}};

\draw [dashed] (3*\widthSim+2*\widthIng+\widthCollect,-\overhead) rectangle (3*\widthSim+2*\widthIng+2*\widthCollect,2*\height+\overhead);

\draw\colorSim (2*\widthSim+2*\widthIng+\widthCollect,\height) rectangle (3*\widthSim+2*\widthIng+\widthCollect,2*\height) node[pos=.5, font=\fontsize{07}{0}\selectfont] {\simu{3}};

 \node[red] (IA2) at (2*\widthSim+2*\widthIng+\widthCollect+\widthRead+\widthAna+\widthSend+\widthIdleA,0.5*\height) {\idle{2}{A}};

\draw\colorTransfer  (3*\widthSim+2*\widthIng+2*\widthCollect,\height) rectangle (3*\widthSim+3*\widthIng+2*\widthCollect,2*\height) node[pos=.45, rotate=-90, font=\fontsize{05.5}{0}\selectfont] {\ingest{3}};

  \draw\colorTransfer  (3*\widthSim+3*\widthIng+2*\widthCollect, 0) rectangle (3*\widthSim+3*\widthIng+2*\widthCollect+\widthRead,\height) node[pos=.5, font=\fontsize{07}{0}\selectfont] {\get{3}};
  \draw\colorAna (3*\widthSim+3*\widthIng+2*\widthCollect+\widthRead,0) rectangle (3*\widthSim+3*\widthIng+2*\widthCollect+\widthRead+\widthAna,\height) node[pos=.5, font=\fontsize{07}{0}\selectfont] {\ana{3}};
 \draw\colorTransfer  (3*\widthSim+3*\widthIng+2*\widthCollect+\widthRead+\widthAna,0) rectangle (3*\widthSim+3*\widthIng+2*\widthCollect+\widthRead+\widthAna+\widthSend,\height) node[pos=.52,rotate=-90, font=\fontsize{05}{0}\selectfont] {\send{3}};

 \draw\colorTransfer  (4*\widthSim+3*\widthIng+2*\widthCollect,\height) rectangle (4*\widthSim+3*\widthIng+3*\widthCollect,2*\height) node[pos=.52, rotate=-90, font=\fontsize{05}{0}\selectfont] {\collect{3}};

 \draw [dashed] (4*\widthSim+3*\widthIng+2*\widthCollect,-\overhead) rectangle (4*\widthSim+3*\widthIng+3*\widthCollect,2*\height+\overhead);

\draw\colorSim (3*\widthSim+3*\widthIng+2*\widthCollect,\height) rectangle (4*\widthSim+3*\widthIng+2*\widthCollect,2*\height) node[pos=.5, font=\fontsize{07}{0}\selectfont] {\simu{4}};

 \node[red] (IA3) at (3*\widthSim+3*\widthIng+2*\widthCollect+\widthRead+\widthAna+\widthSend+\widthIdleA,0.5*\height) {\idle{3}{A}};

\end{tikzpicture} }
   \label{fig:idleA}
}
\caption{Two execution scenarios for the \emd \is workflow.}
  \label{fig:model}
\end{figure}
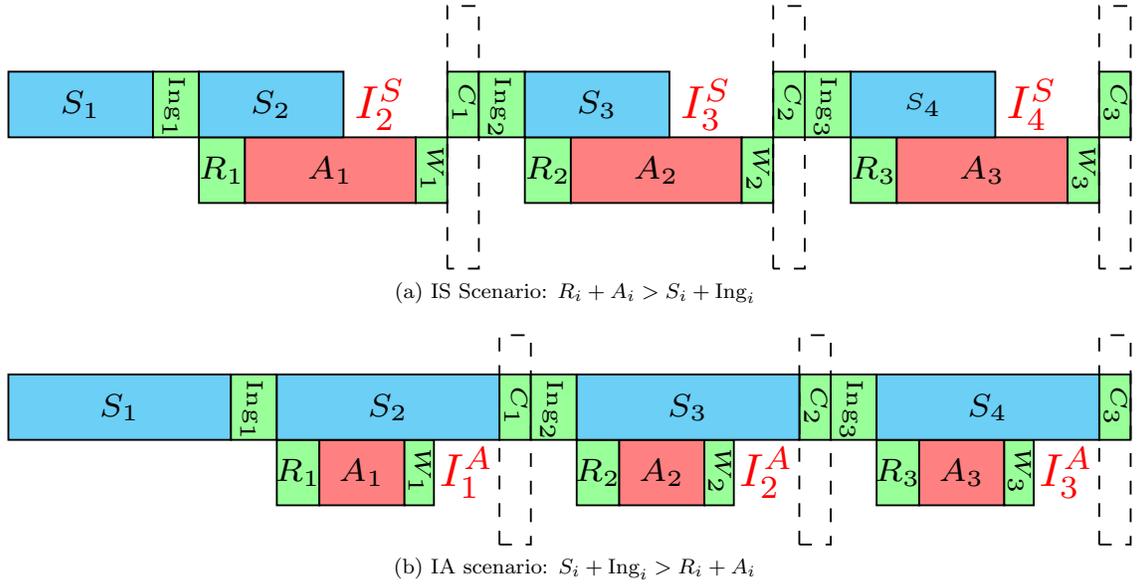

\pp{Workflow Stages and Notations}
Figure~\ref{fig:model} shows the different
stages of the execution of the \emd \is workflow. The simulation component
(\simu{}) produces the data and ingests it into the DTL (\ingest{}). Then, the
analytics component (\ana{}) retrieves this data from the DTL (\get{}) and
processes it. Once done, the analytics component sends the analysis results
back to the simulation component (\send{}), which collects (\collect{}) them
before proceeding with its execution.

This sequence of stages, representing an execution $step$, is repeated
iteratively until the completion of the application. Then, we respectively
denote as \simu{i}, \ingest{i}, \get{i}, \ana{i}, \send{i}, and \collect{i},
the simulation, ingestion, retrieving, analytics, sending, and collection stages
at step $i$. In this model, \simu{i} actually corresponds to the execution of a
certain number of iterations of the main loop of \emd between two analyses. As
mentioned earlier, this number of iterations is denoted as the {\it stride}
which determines how frequently the data analytics component is called.  Then,
for a total number of iterations of the MD simulation loop \iter and a given
stride \stride, the number of steps in our model will be denoted by
$\totalana=\frac{\iter}{\stride}$, and we thus have
$\sum_{i=1}^{\totalana} \simu{i} = \simu {}$ (the same applies for the other
stages). Finally, we make the hypothesis that the different stages are
consistent across steps. This means that the time to execute each stage remains
constant over all the \totalana steps of the application. This hypothesis holds
when the number of steps is large enough ($\totalana \geq 3$), and the impact
of warming-up steps is negligible~\cite{Do21}.

\pp{Dependencies Between Stages}
The data flow within a given step imposes the
following precedence constraints among the different stages:
\begin{equation}
\label{eq:const}
\simu{i} \rightarrow \ingest{i} \rightarrow  \get{i} \rightarrow \ana{i}
\rightarrow \send{i} \rightarrow  \collect{i}\mathrm{,~for~} 1\leq i\leq\totalana.
\end{equation}

Our model also enforces that an analytics stage cannot begin before the
simulation component has received the results of the previous analytics
phase. This means that, from the second step, the ingestion \ingest{i} can be
done if and only if the collection of the metrics from the previous step
\collect{i-1} by the simulation component has been done:
\begin{equation}
\label{eq:synch}
\collect{i-1} \rightarrow \ingest{i}\mathrm{,~for~} 2\leq i \leq \totalana.
\end{equation}

\pp{Idle Periods}
Due to the dependencies between stages described above, idle
periods can appear in the execution flow of the application if one of the
two main components has to wait for the other. Figure~\ref{fig:model}
illustrates two execution scenarios, {\it Idle Analytics} and {\it Idle
  Simulation} in which some idle time occurs. The former (\idleA) corresponds
to a case where the execution time of the simulation related stages (\ie,
simulation and data ingestion) is greater than that of the analytics related
stages (\ie data retrieving and analysis), that is
$\simu{i}+\ingest {i} > \get{i} + \ana{i}$. Conversely, the later (\idleS)
corresponds to analytics stages that take more time than the simulation stages,
\ie $\get{i} + \ana{i} > \simu{i}+\ingest{i}$. We denote by \idle{i}{S}
(resp. \idle{i}{A}) the corresponding idle time for the simulation
(resp. analytics) component at step $i$. Then, we reformulate
Equation~\ref{eq:const} to include these potential idle times:
\begin{equation}
\label{eq:constIdle}
\simu{i} \rightarrow \idle{i}{S} \rightarrow\ingest{i} \rightarrow  \get{i}
\rightarrow \ana{i} \rightarrow \send{i} \rightarrow \idle{i}{A} \rightarrow \collect{i}
\end{equation}

To simplify the notations, we introduce $\simu{*}=\simu{i}$, for all
$i\leq\totalana$. A similar simplification is applied to the other stages in
Equation~\ref{eq:constIdle}.

Assuming that the cost of \collect{*} and \send{*} can be neglected because of
the size of the exchanged data, \ie we consider these stages as synchronization
points, and using the constraints expressed in Equation~\ref{eq:synch}
and~\ref{eq:constIdle}, we define the total idle time \idle{*}{} of a step as:
\begin{equation}
\resizebox{.9\hsize}{!}{
$\idle{*}{} = \idle{*}{S} + \idle{*}{A} =
    \begin{cases}
      \simu{i} + \ingest{i} - (\get{i} + \ana{i}) & \text{if}~\idle{*}{S}=0\\
      \get{i} + \ana{i} - (\simu{i} + \ingest{i}) & \text{if}~\idle{*}{A}=0\\
    \end{cases} = \lvert\simu{i}+\ingest{i}-(\get{i} + \ana{i})\rvert
    $
    }
\end{equation}

Then, an efficient execution of an \is workflow does not induce any idle time,
hence  $\idle{*}{S}=\idle{*}{A}=0$. In other words, such an {\it idle-free}
execution is obtained when $\simu{*}+\ingest{*} = \get{*} + \ana{*}$.

\pp{In-Situ Workflow Execution Efficiency Metric}
Finally, we use the model above to refine a metric presented in~\cite{Do21}
to evaluate the performance of \is workflow executions.
To this end, we define the makespan \makespan{} of the \is workflow as the sum
of the makespan of each step $i\leq\totalana$:
\[ \makespan{} = \sum_{i=1}^{\totalana} \makespan{i}, \]
where \makespan{i} is the maximum of the makespan of each component. Then we
have:
\begin{equation}
\label{eq:makespan}
\makespan{} = \totalana \times \max\left(\simu{*} + \ingest{*}, \get{*} + \ana{*}\right)
\end{equation}

Using this makespan definition, we compute an efficiency ratio \efficiency that
depends on the total idle time induced during the workflow execution. This
ratio is defined as:

\begin{equation}
\label{eq:efficiency}
\efficiency = 1 - \frac{\totalana \times \idle{*}{}}{\makespan{}} =
  1 - \frac{\lvert\simu{i}+\ingest{i}-(\get{i} + \ana{i})\rvert}{\max\left(\simu{*} + \ingest{*}, \get{*} + \ana{*}\right)}
\end{equation}

\subsection{Assessing the Efficiency of In Situ Workflow Executions}

For this evaluation, we consider a 3D Lennard-Jones melt on a
$70 \times 70 \times 70$ region problem instance. The main MD simulation loop
has 8,000 iterations. A performance analysis of \emd allowed us to set the
compute cost per particle to $7.93e$-$7$ and the data size per particle to
transfer to 100.

For the allocation and mapping strategies, we base our study on the work of
Malakar {\it et al.}~\cite{Malakar18} in which they set the simulation to
analysis core allocation ratio \coreRatio as the number of cores allocated to
the simulation component over the number of cores allocated to the data
analytics components. As our target cluster has 32 cores per node, we consider
5 values for this ratio as shown in Table~\ref{tab:coreRatio}. Then, we run
simulations for 1, 2, 4, and 8 nodes (\ie 32, 64, 128, and 256 cores).  All the
results presented hereafter were obtained by running simulations on a single
core of the Dahu cluster.

\renewcommand{\arraystretch}{1.07}
\begin{table}[hbtp]
  \centering
   \caption{Considered simulation to analysis core allocation ratios.}
  \label{tab:coreRatio}
  \begin{tabular}{|c|c|c|} \hline
    {\ema{\bf R}} & {\bf \# simulation cores} & {\bf \# analysis cores} \\
    \hline\hline
    1  & 16 & 16 \\\hline
    3  & 24 & 8 \\\hline
    7  & 28 & 4 \\\hline
    15 & 30 & 2 \\\hline
    31 & 31 & 1 \\\hline
  \end{tabular}
\end{table}

\pp{Impact of Simulation to Analysis Core Allocation Ratio} Let's consider a
user who would like to perform a constant amount of analysis during the
execution of their workflow and know, for a given number of cores, what would
be the most efficient simulation to analysis core allocation ratio to use. This
user can act on two parameters to execute the desired amount of analysis: the
{\it stride} and the {\it cost} of one execution of the data analytics
component. For instance, if the MD simulation loop is executed 8,000 times and
400 units of analysis have to be performed, (at least) four (stride,~cost)
configurations can be envisioned: (20,~1), (200,~10), (500,~25), and (1000,~50).
Thanks to the flexibility of \simsitu, generating a version of the \emd \is
workflow with a larger analysis cost simply amounts to changing one of the
command line parameter, \ie the computing scaling factor.

\begin{figure}[h!]
  \centering
  \includegraphics[width=.95\linewidth]{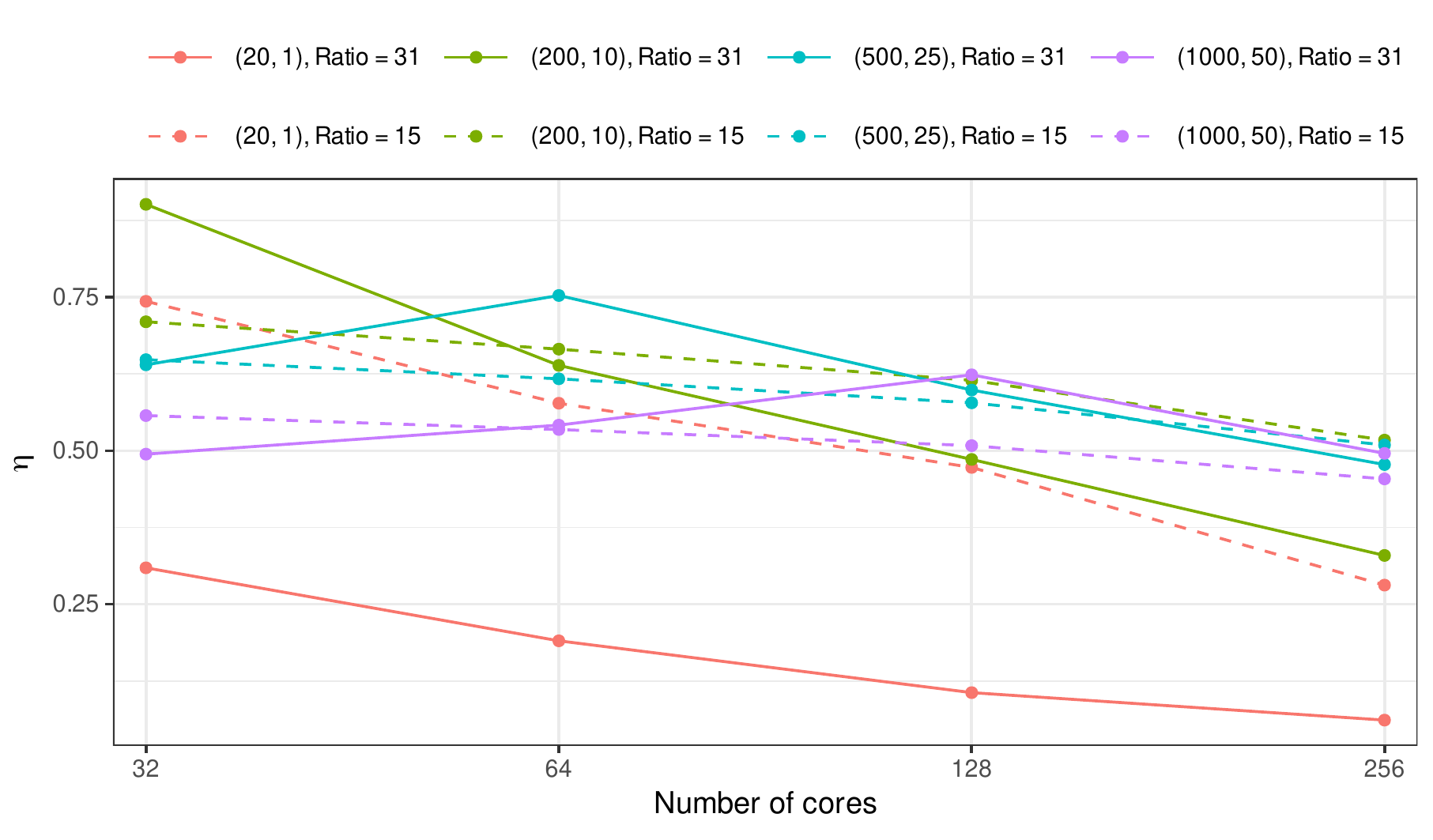}
  \caption{Efficiency of \emd \is workflow in four (stride, analytics cost)
    configurations for two the core allocation ratios.}
  \label{fig:efficiency_100}
\end{figure}

Figure~\ref{fig:efficiency_100} shows the achieved efficiency for these four
combinations of stride and analysis costs and two core-allocation ratios
($\coreRatio = 15$ and $\coreRatio = 31$).  The other ratios lead to lower
efficiency for any core count and are thus not displayed for the sake of
readability. It shows some interesting trends and tradeoffs.  We can see that
the (stride,~cost) configuration that leads to the best efficiency is not the
same for every core count. As the number of cores available for the execution
of the \is workflow increases, it appears to be more efficient to reduce the
frequency and increase the cost of the analytics component. This is confirmed
by the trends of the (20,~1) and (200,~10) configurations with
$\coreRatio = 31$ whose efficiency steadily decreases with the increase of the
number of cores. For these configurations, the analytics actors do not have
enough work to process and thus are idle most of the time. As the total core
count grows, more analytics actors are started, hence amplifying this
phenomenon. A similar trend can be seen for larger analytics cost, but the
tipping point where the efficiency starts to drop is for larger core counts.

We also observe a generally decreasing trend for the efficiency as the number
of cores grows with $\coreRatio = 15$, but over a narrower range. Moreover,
the (200,~10) configuration appears to be consistently achieving good efficiency
for this core allocation ratio, for all total core counts. This better
stability might be preferred by users when they have to adapt their executions
to the number of currently available cores they have access to and do want to
risk to loose efficiency by selecting the wrong configuration.

\begin{figure}[h]
  \centering
  \includegraphics[width=\linewidth]{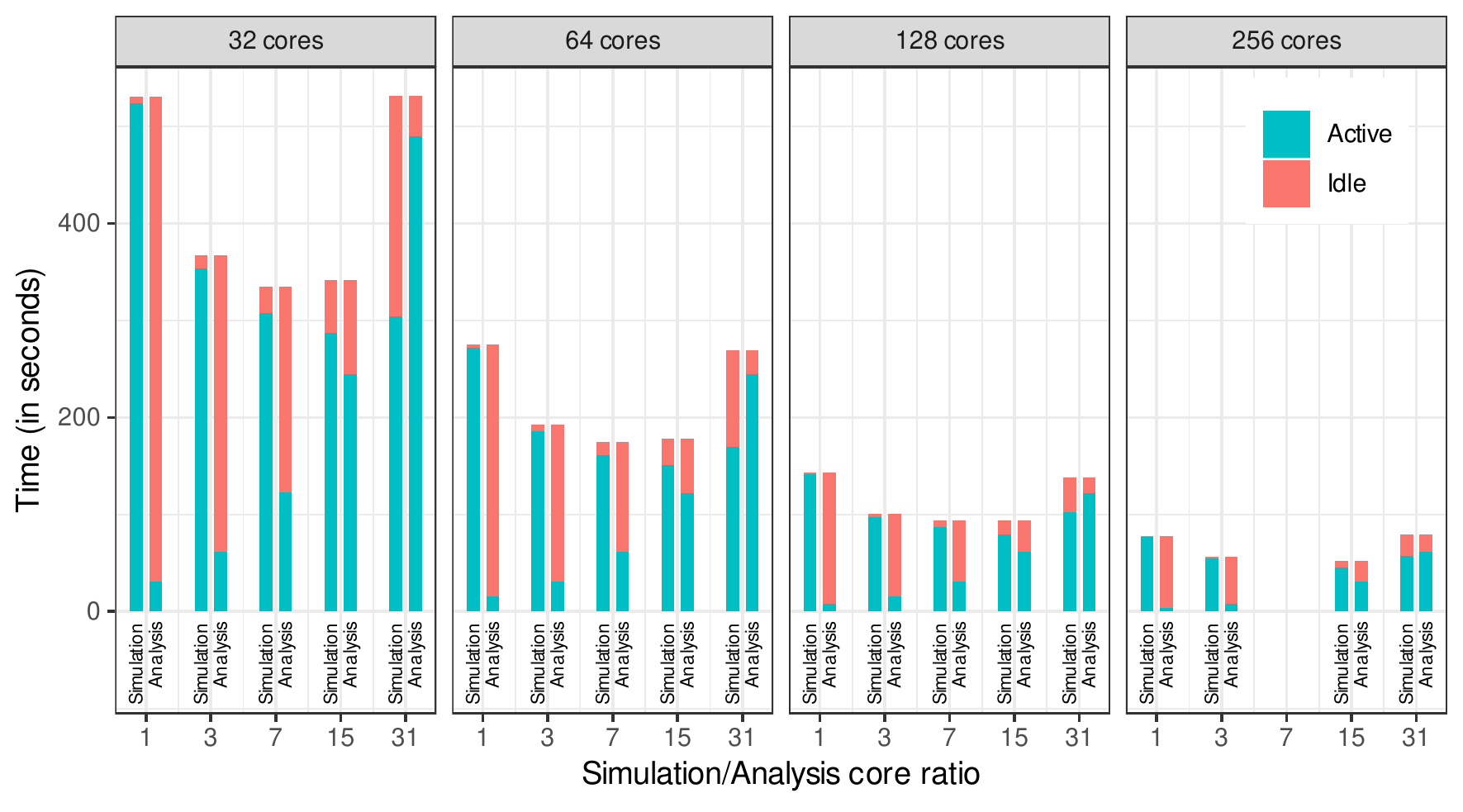}
  \caption{Evolution of the active and idle times of the simulation and
    analytics components when increasing the core allocation ratio and the total
    number of cores for the (1000, 50) scenario.}
  \label{fig:core-ratio-1000-50-mailbox}
\end{figure}

Figure~\ref{fig:core-ratio-1000-50-mailbox} shows a different view of the
(1000,~50) configuration, \ie the evolution of the active and idle times of the
simulation and analytics components when the core allocation ratio and the
total number of cores increase\footnote{The lack of values for 256 cores and
  \coreRatio = 7 is due to a crash of \emd for this particular instance with 224
  MPI ranks (w/ or w/o \simsitu). It seems to come from a badly handled
  division by 0 in \emd's code.}.

In this scenario, we can see that the respective active times of the simulation
and analytics component follow opposite trends. For small core allocation
ratios, the execution time is completely dominated by the time to execute the
MD simulation. Then, the time to execute the analytics component increases
linearly with the ratio, as less cores are allocated to execute the same amount
of analysis, until a tipping point is reached where the simulation waits for
analytics ($\coreRatio=31$).

We also see that some ``sweet spots" can be found, typically for
$\coreRatio=15$, where the active times of the simulation and analytics
components are both efficient and well balanced.  It is interesting to note
that it is in contradiction with the efficiency metric that indicates a better
efficiency with a core allocation ratio of 31 from 64 cores.

These results illustrate how \simsitu can be used to determine a good
core-allocation ratio for a given cost of analysis and number of nodes and that
it is important for users of the \simsitu framework to leverage all the
metrics the tool can provide them while designing their workflow.

\pp{Impact of simulation to analysis data transfers} Another design choice
faced by users of \is workflows is to decide of the mapping of the resources
allocated to the data analytics component. In other words, the question is:
``would it better to adopt an \is strategy, \ie mapping the analytics resources
on the same nodes as the simulation resources, or an \intransit strategy, \ie
dedicating some node(s) to the analytics?''. The former has the advantage of
minimizing the cost of data exchanges between simulation and analytics thanks
to a shared memory space, but the scattering of the analytics resources over
multiple nodes may hinder the performance of this component (\eg communication
intensive analytics function). Conversely, the latter benefits of having all
the analytics located on a single, or small number of dedicated nodes, but
induces a larger communication overhead to exchange data with the simulation
component.

To illustrate how \simsitu can help users to evaluate the relative performance
of \is and \intransit mapping schemes, we consider the following scenario. The
simulation component still corresponds to the main loop of \emd. The analytics
component now corresponds to a function that involves all the analytics
resources and whose performance is impacted by the number of nodes onto which
these resources are allocated, \ie its execution time increases with the
resource scattering. Finally, the user can decide of the volume of data
produced by the simulation to transfer to the analytics.

\begin{figure}[hbtp]
  \centering
  \includegraphics[width=.95\linewidth]{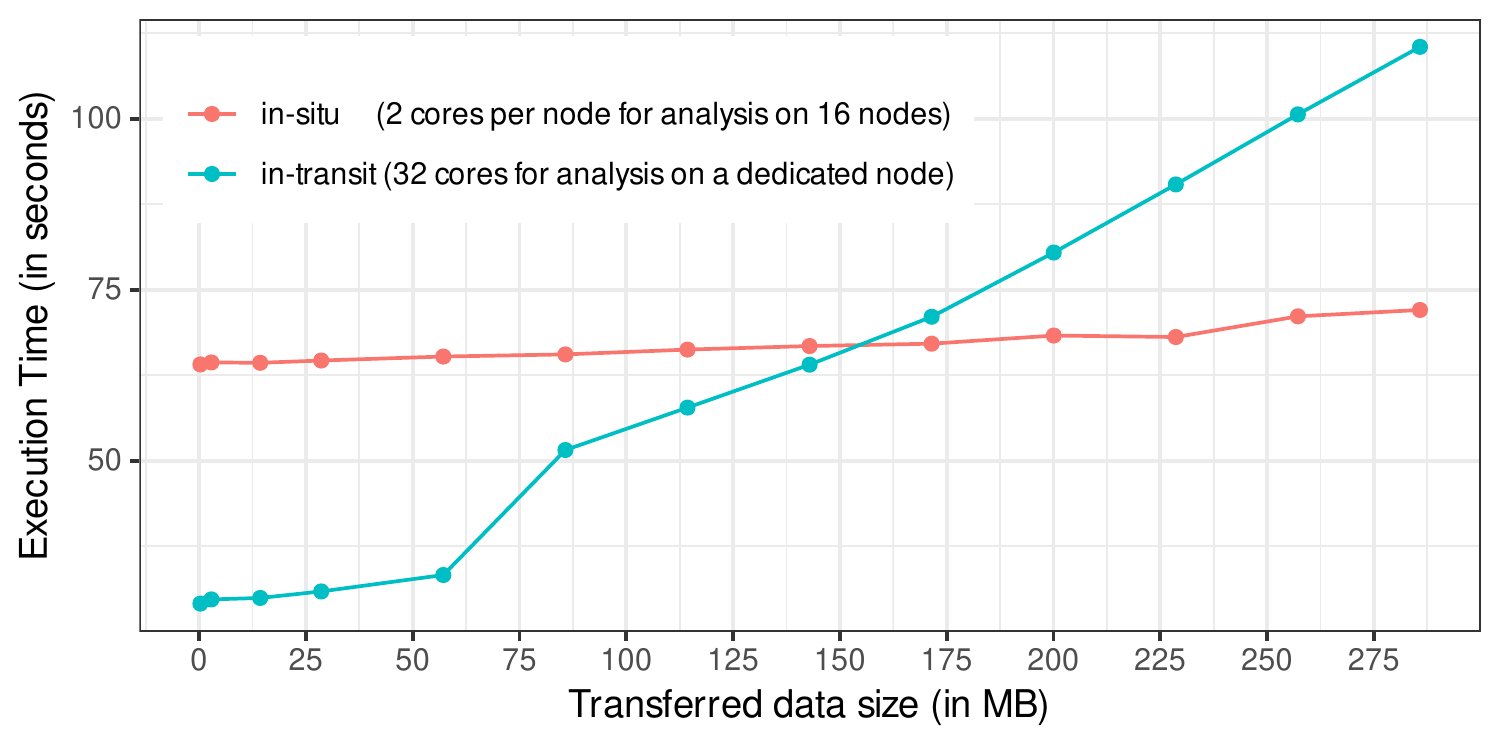}
  \caption{Evolution of the execution time of the simulation component when the
    volume of data transfer is scaled up in the \is and \intransit execution
    modes with 16 nodes and \coreRatio = 15.}
  \label{fig:data_scaling}
\end{figure}

Simulating such a performance study is made easy by the features of
\simsitu. Switching from \is to \intransit mappings simply amounts to change
the analytics hostfile, while changing the volume of transferred data or the
performance profile of the analytics function can be done on command
line. Figure~\ref{fig:data_scaling} shows the evolution of the execution time
of the simulation component of the workflow when the volume of data to exchange
with the analytics component is scaled up to a thousand times. The workflow is
executed on 16 nodes and two execution modes are considered. The \is mode uses
a core allocation ratio of 15, \ie two cores per node are allocated to the
analytics component while a full node is dedicated to the analysis in the
\intransit mode.

We can see that the scattering of the analytics resources across nodes makes
the \is execution mode less efficient than the \intransit execution mode when a
small amount of data is exchanged. However, as we increase this volume of
transferred data, the execution time of the \intransit mode starts to increase
linearly, while the \is execution mode is only slightly impacted as the data
exchanges are done through a shared memory space. Such simulation-based studies
may help users in the design of their analytics component by showing them where
lies the tipping point between \intransit and \is modes for a given
configuration of their workflow.
 \section{Conclusion and Future Work}
\label{sec:ccl}

Analyzing the results of large-scale numerical simulations as they are produced
is an appealing alternative to classical {\it post-hoc} analyses that are more
and more impacted by the increasing discrepancy between the relative
performance of computing and storage subsystems in extreme-scale
supercomputers. However, the development of such \is scientific workflows
raises several challenging questions, such as ``what {\it amount} of analysis
can be done and at which {\it frequency}?", ``how many {\it resources} can
be taken off of the execution of the simulation to execute the analysis?", or
``Is it better to perform \is or \intransit analytics?". 

Determining answers to these questions that do not cause the performance of an
\is workflow to be worse than the classical ``simulation then analysis"
approach usually falls down to evaluating the performance of different
allocation and mapping strategies for different input configurations. However,
the state-of-the-art on the performance evaluation of \is workflows shows that
it relies either on time- and resource-consuming experiments on a limited set
of scenarios or on the simulation of abstracted versions of the initial
applications that may lack of realism.

In this paper, we introduced the \simsitu framework, a generic simulation
framework for \is workflows based on the popular \simgrid toolkit. The modular
design of \simsitu faithfully captures the structure of classical \is workflows
and the behavior of their different components.  We illustrated its simulation
capacities on the \emd Molecular Dynamics proxy-application.  With only a few
minor code modifications, we showed how \simsitu could be used to study
different execution scenarios of \is workflows and highlight important
performance tradeoffs.

As part of our future work, we plan to further demonstrate the simulation
capacities of \simsitu by investigation more allocation and mapping strategies
on different use-case applications. We will particularly focus on \intransit
processing where nodes are dedicated to analytics. Studying such strategies
would be a first step in evaluating the impact of data transfers and network
performance on the execution of scientific workflows. We also plan to extend
the simulation capacities and realism of \simsitu by developing more complex
versions of the Data Transport Layer component that mimic the behavior of
popular implementations such as DataSpaces~\cite{dataspaces} or Dimes~\cite
{dimes}. The objective is to provide \simsitu users with the capacity to select
which flavor of the DTL they want to use in the simulation of their \is
workflows. Finally, we plan to leverage \simsitu to carry out performance
evaluations in scenarios that would be hardly possible to evaluate through
actual experiments on supercomputers. For instance, the necessity of running
series of parallel MD simulations in ensembles~\cite
{chelli2012ensemble} broadens the range of feasible \is configurations and
raises new allocation and scheduling challenges. Moreover, the modularity of
the \simsitu framework offers enough flexibility to envision the online
evaluation of scheduling decisions in the context of an adaptive sampling
process~\cite{hruska2020adaptive}.
 
{
\small
\medskip
\noindent
\textbf{\emph{Acknowledgments.}}
Experiments presented in this paper were carried out using the Grid'5000
testbed, supported by a scientific interest group hosted by Inria and including
CNRS, RENATER and several Universities as well as other organizations (see
\url{https://www.grid5000.fr}). This research used resources of the Oak Ridge Leadership Computing Facility 
at the Oak Ridge National Laboratory, which is supported by the Office of 
Science of the U.S. Department of Energy under Contract No. DE-AC05-00OR22725.
}

\balance
\bibliographystyle{IEEEtran}
\bibliography{biblio}

\end{document}